\documentclass[showkeys,twocolumn,preprintnumbers,amsmath,amssymb,amsfonts,amsthm]{revtex4}
\usepackage{amsthm}% Include figure files
\usepackage{dcolumn}% Align table columns on decimal point
\usepackage{bm}% bold math

\newcommand{\tr}{{\rm tr}}
\newtheorem{definition}{Definition}[section]
\newtheorem{thm}{Theorem}[section]
\newtheorem{cor}[thm]{Corollary}
\newtheorem{prop}{Proposition}[section]

\theoremstyle{remark}

\theoremstyle{plain}

\newtheoremstyle{note}% name
  {3pt}%      Space above
  {3pt}%      Space below
  {}%         Body font
  {}%         Indent amount (empty = no indent, \parindent = para indent)
  {\itshape}% Thm head font
  {:}%        Punctuation after thm head
  {.5em}%     Space after thm head: " " = normal interword space;
  {}%         Thm head spec (can be left empty, meaning `normal')

\theoremstyle{note}

\newtheoremstyle{citing}% name
  {3pt}%      Space above, empty = `usual value'
  {3pt}%      Space below
  {\itshape}% Body font
  {}%         Indent amount (empty = no indent, \parindent = para indent)
  {\bfseries}% Thm head font
  {.}%        Punctuation after thm head
  {.5em}%     Space after thm head: " " = normal interword space;
  {\thmnote{#3}}% Thm head spec

\theoremstyle{citing}
\newtheoremstyle{break}% name
  {9pt}%      Space above, empty = `usual value'
  {9pt}%      Space below
  {\itshape}% Body font
  {}%         Indent amount (empty = no indent, \parindent = para indent)
  {\bfseries}% Thm head font
  {.}%        Punctuation after thm head
  {\newline}% Space after thm head: \newline = linebreak
  {}%         Thm head spec

\theoremstyle{break}

\theoremstyle{exercise}

\swapnumbers
\theoremstyle{plain}

\let\lvert=|\let\rvert=|

\begin{document}

%\preprint{APS/123-QED}

\title{A Fixed Point Result for Environment-Induced Semigroups}% Force line breaks with \\

\author{Andreas Martin Lisewski}
% \altaffiliation[Also at ]{Belgradstrasse 19, 80796 Munich, Germany}%Lines break automatically or can be forced with \\
 \email{martin.lisewski@onlinehome.de}
\affiliation{%
Belgradstrasse 19, 80796 Munich, Germany
}%


\begin{abstract}
Based on the environment-induced semigroup approach to the quantum measurement process, we show that a certain class of these semigroups, referred to as contractive uniformly $k$-Lipschitzian semigroups, exhibit a fixed point property. With regard to the quantum measurement problem, semigroups of this kind ensure decoherence and the selection of a single state from the familiy of pointer states. In fact, the common fixed point is the selected state.  
\end{abstract}

%Use showkeys class option if keyword
\maketitle

\section{\label{Introduction}Introduction}

The theory of quantum measurement aims to explain the emergence of certain objective results out of an originally given physical structure that consists of three entities: a quantum physical system that is object to a measurement (\emph{the system}, in short), the physical environment of this system (\emph{the environment}), and a human subject. The system itself is any part of nature where the laws of quantum mechanics may reasonably be applied to, while the environment is the physical complement of the system consisting of the measurement apparatus as well as of all physical structures that exert some non neglegible influence to the system. And finally, the human subject being a conscious witness to the experimental outcome of the measurement and treating this outcome as an objective document.    It turns out that under suitable conditions a measurement result may properly be represented as a unique classical quantity. Now this situation poses a problem,  because before measurement this quantity did not appear to play a special or an outstanding role within the quantum physical description of the system nor did this quantity have a special meaning in the physical description of the system's environment including the human observer. More precisely, even though quantum theory permits a variety  of different experimental outcomes (allowing also for non classical results, i.e., quantum superpositions), a measurement of a quantum system is often characterized by a single outcome that corresponds to a classical state. Therefore a theory of quantum measurement must be able to explain at least two physical processes:
\begin{enumerate}
\item The effective elimination of all quantum superpositions, i.e.,  states of the kind ''Schr\"odinger's cat''.
\item The selection of a single state out of the set of all remaining states.  
\end{enumerate}  

Several theoretical attempts have been made to accomplish these objectives, i.e., to explain the apparent collapse of the wave function. In fact, one particular method--referred to as decoherence-- proved to be successful in explaining the absence of certain quantum states in measurement results. With decoherence it has been shown that the interaction between the measured quantum system and its environment--in particular the measurement apparatus--  eliminates exactly those quantum superpositions which according to classical physics represent contradictory facts, e.g., a radioactive nucleus being in a state after and before its decay at the same time. Thus decoherence is able to explain the first of the two porcessess stated above. However, with regard to the second point a convincing answer within the decoherence framework has not been given yet. For example, recent arguments as given by Adler \cite{a2002} shed doubts on the ability of the decoherence approach to select a single quasi-classical outcome of a quantum measurement. These arguments seem to invalidate an opinion which according to decoherence can indeed accomplish the first \emph{and} the second objective-- as has been put forward recently by several authors (c.f., references in \cite{a2002}).  

The aim of this text is to show that there exists a plausible model of decoherence that can indeed accomplish both objectives. We follow the seminal works of Olkiewicz and Blanchard \cite{bo2000}, \cite{o2000}, and approach the quantum measurement problem by treating environment (essentially, the measurement apparatus) as a classical system described by a commutative algebra of functions. In this approach, the evolution equation of the classical part is modified by the expectation value of some quantum observable of the quantum system while, at the same time, the Schr\"odinger unitary dynamics for the latter is replaced by a \emph{environment-induced semigroup} $\mathcal{T}$ of positive maps (Henceforth, we will frequently use the abbrevation EIS.). This semigroup is generated by a Markovian master equation. As a consequence, the evolution of the quantum system becomes dissipative and (Markovian) stochastic through the interaction with its environment.
This method represents decoherence as a process of continuous measurement-like intercation between the system and its environment in which the environent-induced semigroup  naturally leads to a decomposition of the space of observables: one subspace in which the semigroup acts in a reversible unitary way (referred to as the isometric subspace), and the second in which the semigroup sweeps out the rest of the statistical states (referred to as the sweeping subspace). The isometric part is used to define sets of \emph{classical states}. If they turn out to be nonempty, Blanchard and Olkiewicz conclude that these states correspond to points in a classical phase space. 

With the environment-induced semigroup approach, we look at a system's transition in time from an initial phase where the system is described completely by means of unitary Schr\"odinger dynamics (the closed phase), to  a secondary phase where the measurement-like interaction between the system and its environment is described by means of the EIS (the open phase). Our main task is to demonstrate that under some general conditions the dynamics in the open phase is characterized by an attractive fixed point of the EIS in the quantum state space, and that this point corresponds to a point in classical phase space. Thus we conclude that if decoherence is represented by an environment-induced semigroup, it can accomplish the first and the second objective (as stated above) that every physical model of quantum measurement has to comply with.

%============================

\section{The Fixed Point}
\subsection{``Classical'' States}

Environment-induced semigroups form  a subset of so-called dynamical semigroups, i.e., a strongly continuous semigroup of completely positive trace preserving and contractive (in the trace norm $\|\cdot\|_1$) operators acting on the space of trace class operators $ \mathcal{C}_1$ (This Banach space is often referred to as the von Neumann-Schatten class.). In addition, it is also contractive in the operator norm $ \| \cdot \|_\infty$; this last property ensures that for an evolving  statistical operator $ \rho$ the statistical entropy $ S(\rho) = -\tr \rho \log\rho$ and the linear entropy (often referred to as the purity) $ S_l(\rho) =1 -  \tr\rho^2$ never decrease \cite{o2000}. Also, an environment-induced semigroup $ \mathcal{T} = \{T_t:t\geq0\}$, determines two linear, closed  and $\mathcal{T}$-invariant  subspaces $\mathcal{C}_1^{\rm i}$ and $ \mathcal{C}_1^{\rm s}$ in the Banach space of all trace class operators $ \mathcal{C}_1$. The subspace $\mathcal{C}_1^{\rm i}$ is called the isometric part and $\mathcal{C}_1^{\rm s}$ the sweeping part. These spaces have the following properties \cite{o1999}:\\
a) $\mathcal{C}^{\rm i}_1$ and $\mathcal{C}^{\rm s}_1$ are *-invariant;\\
b) For all $e_1 \in \mathcal{C}^{\rm i}_1, e_2 \in \mathcal{C}^{\rm s}_1$ it is $ \tr[e_1e_2] = 0$;\\
c) $\mathcal{C}_1 = \mathcal{C}^{\rm i}_1 \oplus \mathcal{C}^{\rm s}_1$;\\
d) $T_t\arrowvert_{\mathcal{C}^{\rm i}_1}$ is an invertible isometry given by a unitary group, i.e., $T_t\arrowvert_{\mathcal{C}^{\rm i}_1} e_1 = U_te_1U^*_t$ for any $ e_1 \in \mathcal{C}^{\rm i}_1$;\\
e) weak*$-\lim_{t\rightarrow\infty}T_t\arrowvert_{\mathcal{C}^{\rm s}_1}e_2 = 0$ in the weak* topology for any $ e_2 \in \mathcal{C}^{\rm s}_1$.

 It can be shown that any one-dimensional projection $ e \in \mathcal{C}_1^{\rm i}$ remains a projection during the temporal evolution, and so $ S_l(T_te) = 0$ , $ T_t \in \mathcal{T}$, for all $ t \geq 0$ (In other words, the temporal evolution in the isometric part is fully determined by Schr\"odinger dynamics.). This property may be used to define the set $ \mathcal{S}_0$ of ``robust states''
\begin{equation}
\mathcal{S}_0 = \mathcal{S} \cap \mathcal{C}_1^{\rm i}\,,
\end{equation}
where $ \mathcal{S}$ is the set of all states. By a state we always mean a pure state; hence $ \mathcal{S}$ consists of one-dimensional projections (determined up to a complex phase factor) in a Hilbert space $ \mathcal{H}$. An arbitrary  state from $ \mathcal{S}_0$ will remain robust, i.e., will still be an element of  $ \mathcal{S}_0$  and thus will stay pure during time evolution. In other words, robust states evolve unitarily. Thus the set of robust states is likely to contain ``classical'' states, because unitary evolution guarantees perfect predictability-- surely a desired feature of any state attributed as being ``classical''.  However, the isometric subspace may contain states that do not have a classical counterpart. Since the quantum mechanical superposition of states may still be applied, a linear combination of, say, two robust states may result in another robust state. This feature has no classical analogy, since classical (and deterministic) states do not combine into another classical state. Blanchard and Olkiewicz use this observation to state the following definition:
\begin{definition}
A state $ e \in \mathcal{S}$ is called ``classical'' if $ e \in \mathcal{S}_0$ and if for any $ f \in \mathcal{S}_0, \, f \neq e, \, S(e, f) \cap \mathcal{S}_0 = \emptyset$, where $ S(e, f)$ denotes the set of all states being non-trivial superpositions of $e $ and $ f$. The collection of all ``classical'' states is denoted by $ \mathcal{S}_c$.
\end{definition}
It can be shown that although ``classical'' states remain pure under the action of an arbitrary element, $ T_t$, the environment-induced semigroup, any of their non-trivial superpositions loses its purity, i.e. evolves into a mixture. Given that $ \mathcal{S}_c$ is non-empty, the following  statement about its elements can be made \cite{bo2000}: 
\begin{thm}
\label{thm1}
If $ \mathcal{S}_c \neq \emptyset$, then it consists of a family, possibly finite, of pairwise orthogonal states $ \{ e_1, e_2, \dots\}$ such that $ T_t e_i = e_i$ for all $ t \geq 0$ and any index $ i$.  
\end{thm}
This theorem reveals two important features. First, since it is $ e_i \cdot e_j = \delta_{ij}\,e_i$, it is evident that ``classical'' states form a so-called pointer basis, which means that the corresponding reduced density matrix is diagonal in this particular basis. Second, since it is $ T_t e_i = e_i$ for all $ i$ and $t \geq 0$, it turns out that each ``classical'' state is actually a common fixed point of the environment-induced semigroup $\mathcal{T}$.

It is clear that if before measurement the quantum system is in an eigenstate of the measured observable, then this state coincides with a ``classical'' state. But what happens when the quatum system is not in an eigenstate of the measured observable? Will then ``classical'' states emerge anyway? In this context it is worthwhile to note that theorem \ref{thm1} tells nothing about the actual existence of ``classical'' states. Furthermore, and in the view of the statements made in the introduction, a proper process that selects a single state out of the pointer basis of ``classical'' states is still missing. Thus an answer to the question of how a unique and classical measurement result is ever achieved has not been addressed yet. 

\subsection{The Fixed Point}

In this part we extend the methods presented in the previous section, such that we will arrive at a situation that allows us to discuss the open questions stated above.

Here, we presume that the environment-induced semigroup is contractive in the sense that for every $ f,g \in \mathcal{C}_1$ with $f \neq g$ the inequality
\begin{equation}
\label{contract}
\|T_tf - T_tg\|_1 < \|f - g\|_1
\end{equation} 
holds for all $ t \geq 0$. Moreover, if $ f$ is positive with $ \|f\|_1 = 1$ then $\|T_tf\|_1 = \|f\|_1$. 

First of all, we observe that the existence of a ``classical'' state $ e \in \mathcal{S}_c$ is equivalent to the existence of a unique common fixed point $T_te = e$ for all $ t \geq 0$. Since theorem \ref{thm1} tells us $ e \in \mathcal{S}_c \Rightarrow T_t e = e$ for all $ t \geq 0$, it is sufficient to show the opposite implication. Thus let $ e$ be a common fixed point of the environment-induced semigroup $ \mathcal{T} = \{T_t: t \geq 0\}$ acting on the Banach space $\mathcal{C}_1$. Since $T_t$, $ t \geq 0$, is contracting in the operator norm $ \|\cdot\|_1$, there are no other fixed points. Furthermore, we show that $ e$ must be an element of $ \mathcal{S}_0$. It is clear that $ e \in \mathcal{C}_1^{\rm i}$, because obviously an element of the sweeping space $ \mathcal{C}_1^{\rm s}$ cannot be a fixed point of $ T_t$. We then assume that $ e \in \mathcal{C}_1^{\rm i}$ while $ e \notin \mathcal{S}_0$. But this assumption leads to a contradiction. This stems from the fact that an element $ e \in \mathcal{C}_1^{\rm i}\setminus\mathcal{S}_0$ is either not pure, or, if it is a superposition of distinct pure states, its trace norm  $\|\cdot\|_1$ must not equal one. But on the other hand we know that $T_t$ acts on $ \mathcal{C}_1^{\rm i}$ as an invertible isometry given by a unitary group -- c.f., property d) in the previous paragraph, and that the unitary evolution generated by such a group can never change a pure state into a mixture or change the norm of the pure state. Moreover, let the set $ \mathcal{S}_0$ be nonempty, and let $ f \in \mathcal{S}_0$, then our assumption together with the prerequisite that $ T_t$ is contractive in the $ \| \cdot \|_1$ norm demand 
\begin{equation}
\label{lim}
\lim_{t \rightarrow \infty} \| T_te - T_t f \|_1 = \lim_{t \rightarrow \infty} \| e - T_t\arrowvert_{\mathcal{C}^{\rm i}_1} f \|_1 = 0 \,.
\end{equation}
But since $e$ is a mixture or it is $ \|e\|_1 \neq 1$, while $f$ is a pure state, the unitary (and continuous) evolution of $ T_t\arrowvert_{\mathcal{C}^{\rm i}_1}f$ cannot be brought in line with equation (\ref{lim}). Thus we have shown that $ e \in \mathcal{S}_0$. Finally, let again $ S(e,f)$ denote the set of all states being non-trivial superpositions of $e \in \mathcal{S}_0$ and $f \in \mathcal{S}_0, \, e \neq f$. Then, clearly,  $ S(e,f) \cap \mathcal{S}_0  = \emptyset$. So, we have proven the following proposition:
\begin{prop}
\label{prop1}
There is at most one $ e \in \mathcal{C}_1$ such that the following equivalence relation holds: $ e \in \mathcal{S}_c \Leftrightarrow e$ is a unique common fixed point of the environment-induced semigroup $ \mathcal{T}$.
\end{prop}
Our next intention is to investigate the existence of a fixed point. We start with preliminary defnitions. 
Let $ G$ be a semitopological semigroup, i.e., $ G$ is a semigroup woth a Hausdorff topology such that for each $ a \in G$, the mappings $ t \rightarrow at$ and $ t \rightarrow ta$ from $ G$ into itself are continuous. Let $ C$ be a nonempty subset of a Banach space $ X$. Then the familiy $ \mathcal{T} = \{ T_t: t \in G \}$ of self-mappings of $ X$ is said to be a \emph{Lipschitzian semigroup} on $ X$ if the following properties are satisfied \cite{tx95}:
\begin{enumerate}
\item $T_{ts}x = T_tT_sx$ for all $t, s \in G$ and $ x \in X$;
\item for each $ x \in C$ the mapping $ t \rightarrow T_t x$ is continuous on $ G$;
\item for each $ t \in G$ there is a constant $ k(t) > 0$ such that $ \|T_tx - T_ty\| \leq k(t) \|x - y\|$ for all $ x,y \in X$.
\end{enumerate}
A Lipschitzian semigroup $ \mathcal{T}$ is called \emph{uniformly k-Lipschitzian} if $ k(t) = k$ for all $ t \in G$ and in particular, \emph{nonexpansive} if $ k_t = 1$ for all $ t \in G$. It is evident that any environment-induced semigroup becomes a nonexpansive uniformly $k$-Lipschitzian semigroup if we identify $ X$ as the Banach space of trace class operators $ \mathcal{C}_1$ equipped with the trace norm $ \| \cdot \|_1$, and $ G$ as $ \mathbb{R}^+$. In fact, any environment-induced semigroup is even contractive, i.e.,  it is $ k(t) < 1$ for all $ t \geq 0$.

For any $ x \in X$, the orbit of $ x$ under $ \mathcal{T}$ starting at $ x$ is the set
\begin{equation*}
O(x) = \{x\} \cup \{ T_tx: t \in G\} \,,
\end{equation*}
and for any $ x, y \in X$ we set $O(x,y) = O(x) \cup O(y)$. A subset $ C \subset X$ is said to be bounded if its diameter $ {\rm diam}(C)$, defined as 
\begin{equation*}
 {\rm diam}(C) = \sup\{\|x - y \|:x,y \in C\}\,,
\end{equation*}
is finite. Additionally, we say a semigroup $ \mathcal{T}$ is near-commutative if, for any $ t,s \in G$, there exists a $ u \in G$ such that it is $ T_tT_sx = T_sT_ux$ for all $ x \in X$. With these definitions the following theorem holds (c.f., \cite{hh99}):
\begin{thm}
\label{thm2}
Suppose that $ \mathcal{T}$ is a near-commutative semigroup of continuous self-mappings on a Banach space $ X$ such that the two following conditions are satisfied
\begin{enumerate}
\item For any $ x \in X$, its orbit $ O(x)$ is bounded.
\item There exists an upper semicontinuous function $ \varphi:[0, \infty) \rightarrow [0,\infty)$ such that $ \varphi(0) = 0$ and $\varphi(a) < a$ for any $ a > 0$ with the property that, for any $T_t \in \mathcal{T}$, there exists $ n(T_t) \in \mathbb{N}$ such that $\|T_t^nx - T_t^ny\| \leq \varphi({\rm diam}(O(x,y)))$ for all $ n \geq n(T_t)$ and $x,y \in X$.
\end{enumerate}
Then $\mathcal{T}$ has a unique common fixed point $ x^* \in X$ and, moreover, for any $ T_t \in \mathcal{T}$ and any $ x \in X$ it is 
\begin{equation}
\lim_{n \rightarrow\infty}\| T_t^nx - x^*\| = 0. 
\end{equation}
\end{thm}
First we note that any EIS is near-commutative because it is commutative. We presume again that for a given EIS $ \mathcal{T}$ the set of robst states $ \mathcal{S}_0$ is not empty; and we further assume that the environment-induced semigroup is uniformly contractive, which means that $ k = \sup\{k(t): t \geq 0\} < 1$. With this assumption we may construct an upper semicontinuous  ``gauge function'' $\varphi: [0,\infty) \rightarrow [0,\infty)$, viz. $\varphi(a) = k \,a$ for all $ a \geq 0$ (A function $f: [0,\infty) \rightarrow [0,\infty)$ is said to be upper semicontiuous at a point $ x_0$ if for any $ \epsilon > 0$ there is a $ \delta > 0$ such that $f(x) - f(x_0) < \epsilon$  for all $ x \in [0,\infty)$ with $|x - x_0| < \delta$. This function is called upper semicontinuous if it is upper semicontiuous at all points $ x_0 \in [0,\infty)$.). 

Next we verify the two conditions given in theorem \ref{thm2}. Given any $ f \in \mathcal{C}_1$ its orbit $ O(f)$ has to be bounded because for $ g \in \mathcal{S}_0$ it is $ \|T_tg\|_1 = 1$ for all $ t \geq 0$, which means that $ O(g)$ is bounded. At same time it must be $ \|T_tg - T_tf\|_1 < \|g - f\|_1$ for all $ t \geq 0$; thus we have $\|T_tf\|_1  \leq  1 + \|T_tg - T_tf\|_1 < 1 + \|g - f\|_1 \leq 2 + \|f\|_1$. Therefore it is $ {\rm diam}(O(f)) \leq 2(2 + \|f\|_1$) for any $ f \in \mathcal{C}_1$. The second condition is also fulfilled; as it is
\begin{eqnarray*}
\|T^n_tx - T^n_ty\|_1 &=& \|T_{nt}x - T_{nt}y\|_1
 \leq k \|x - y\|_1\\
& = & \varphi(\|x - y\|_1)
 \leq \varphi(O(x,y))
\end{eqnarray*}
for every $ n \in \mathbb{N}$ and $ t \geq 0$. Thus we obtain a fixed point result:
\begin{prop}
\label{prop2}
Any environment-induced semigroup $ \mathcal{T} = \{T_t: t \geq 0\}$, which is uniformly $k$-Lipschitzian on $ \mathcal{C}_1$ with $ k < 1$, and which implies $ \mathcal{S}_0 \neq \emptyset$, has a unique common fixed point in $\mathcal{C}_1$, i.e., there is exactly one $e \in \mathcal{C}_1$ such that $ T_t e = e$ for all $ t \geq 0$. Moreover, for an arbitrary  $ f \in \mathcal{C}_1$ and any $ T_t \in \mathcal{T}$ it is
\begin{equation}
\lim_{t \rightarrow \infty} \| T_tf - e\|_1 = 0 \,.
\end{equation}
\end{prop}
Given explicit existence, we may, by virtue of proposition \ref{prop1}, imply
\begin{cor}
$e $ is ``classical'', i.e., $ e \in \mathcal{S}_c$.
\end{cor}
We remark here that proposition \ref{prop2} does not establish a fixed point property for the whole class of  possible environment-induced semigroups; instead it is valid only for a proper subclass whose members are uniformly contractive $k$-Lipschitzian semigroups. Nevertheless,  this proposition and the following corollary tell us that there is indeed a unique ``classical'' state (among a family of pointer states) such that it may be regarded as the unique and classical outcome of a continuous measurement process represented by an EIS acting on a open quantum system. Furthermore, our result does not determine which state from the pointer basis becomes a common fixed point of the EIS; thus it does not introduce a deterministic measurement process. Also, since all states converge towards the same fixed point within the open phase, the ultimate ($ t \rightarrow \infty$) outcome of the measurement process is independent of the state of the quantum system prior to the measurement, i.e., the state in the closed phase. The fixed point property solely depends on the set of the quantum system's statistical operators  -- embedded in the Banach space $ \mathcal{C}_1$, and on its dynamical interaction with an environment-induced semigroup. This interaction is encoded in the algebraic and analytical structure of  the isometric-sweeping space decomposition: $ \mathcal{C}_1 = \mathcal{C}_1^{\rm i} \oplus \mathcal{C}_1^{\rm s}$.
\section{Concluding Remarks}
We have shown that a certain class of environment-induced semigroups, i.e., that of all contractive uniformly $k$-Lipschitzian semigroups, does indeed accomplish \emph{both} of the desired physical processes that we expect from a physical model of quantum measurement, namely decoherence \emph{and} the selection of a single ``classical'' state out of  the familiy of pointer states. Thus the environment-induced semigroup approach for continuous quantum measurements may well be regarded as a promising physical model. 
\vfill

However, several open questions considering the prerequisites and the consequences of our results remain. We name here a few without going into details. First, one may ask for the conditions where it becomes permissable to represent a continuous measurement as a Markovian master equation. A non-Markovian evolution equation does not generate a continuous semigroup of completely positive maps, i.e., in this case a semigroup composition law $ T_sT_t = T_{st}$ is missing, and so the results presented here do not apply. Second, it is worthwhile to explore necessary conditions that imply a common fixed point property for a larger class of EIS, i.e. for Lipschitzian semigroups where $ k$ is not uniformly smaller than one. Third, with regard to the structure of fixed point itself, one may ask under what circumstances the fixed point becomes ergodic. And fourth, one may investigate the interaction between a given EIS and local group transformations in the quantum phase space. Transformations of this kind are associated with a Lie group $ G$ that represents the non-locality of quantum states. Examples of realizations of this group are: $ G = U(1)\times SU(2)$ for a spin-$1/2$ particle, the cyclic group $ G = \mathbb{Z}_p$ ($ p$ a prime) for a quantum harmonic oscillator, or the Heisenberg-Weyl group $ G = HW(q,p)$ for a free particle. Since in general these groups do not act as contractions on corresponding quantum state spaces, it would be interesting to study possible emerging phenomena such as symmetry breaking.     
\vfill
\bibliography{phasequant_fixed}

\end{document}